\documentclass{PoS}

\usepackage[authoryear]{natbib} 
\bibpunct{(}{)}{;}{a}{}{,}      

\def\gray{$\gamma$-ray\ }
\def\grays{$\gamma$-rays\ }
\def\Etrue{E_{true}}
\def\Emeas{E_{meas}}

\def\Egamma{E_\gamma}
\def\qhad{q_{had}}
\def\qlep{q_{lep}}
\def\qtot{q_{tot}}
\def\fluxunits{cm$^{-2}$~sr$^{-1}$~s$^{-1}$~MeV$^{-1}$\ }
\def\Fermi{{\it{Fermi}}}

\title{Local interstellar cosmic-ray spectra derived from gamma-ray emissivities}

\ShortTitle{Local interstellar cosmic-ray spectra}

\author{\speaker{A. W. Strong}{, on behalf of the \Fermi-LAT Collaboration}\\
        Max-Planck-Institut f\"ur extraterrestrische Physik\\
        E-mail: \email{aws@mpe.mpg.de}}

\abstract{Precise gamma-ray emissivities from cosmic-ray interactions with interstellar gas have been recently derived using \Fermi-LAT data,  and used to constrain the local interstellar spectra of protons and leptons. We report on a continuing effort to exploit these emissivities combined with the latest hadronic gamma-ray production cross-sections and other constraints such as synchrotron emission for the leptonic component. The interstellar spectra provide important information for heliospheric modulation, and cosmic-ray origin and propagation.}

\FullConference{The 34th International Cosmic Ray Conference,\\
		30 July- 6 August, 2015\\
		The Hague, The Netherlands}

\begin{document}
\section{Introduction}

The spectrum of cosmic rays (CR) in the local interstellar medium (within about 1 kpc from the Sun) is of interest as a complement to direct measurements and as a probe of solar modulation which affects particles below a few GeV.
\grays from CR protons and heavier nuclei interacting with interstellar gas are an ideal probe of local CR; bremsstrahlung from CR electrons and positrons is also important at low energies and must be accounted for too. 
In \cite{2013arXiv1307.0497D,2013arXiv1303.6482D} 
 we presented preliminary results based on earlier emissivity data \citet{casandjian2013}.
\cite{2015arXiv150505757D} 
 addresses other aspects of this topic, also using those emissivities.

Recently  a new and precise determination of the local \gray emissivity using \Fermi-LAT data has been made \citet{2015arXiv150600047C}, studying regions at Galactic latitudes out of the plane; in particular the contributions from atomic, molecular and ionized hydrogen were separated,
 and the emissivity of atomic gas traced by the 21-cm line can be used as the most reliable data for analysis. \citet{2015arXiv150600047C} also provided an extensive analysis of the emissivities to derive the interstellar proton spectum.
Here we pursue this project with the new emissivities, with various innovations including new cross-sections and  analysis techniques.

\section{Emissivity matrix representation}
\subsection{Response Matrices}
The observed emissivity is the sum of hadronic (pion-production) and leptonic (bremsstrahlung) contributions:
$\qtot(\Egamma)=\qhad(\Egamma)+\qlep(\Egamma)$.
The gamma-ray emissivity  predicted for  the input CR spectra can be written in matrix form
$\epsilon_k^{pred} = \Sigma_{process}\Sigma_j M_{jk}I_j(\vec\theta)$.
where $\epsilon_k$ is the emissivity in the k'th energy bin, and $I_j(\vec\theta)$ is the j'th momentum sample point of the CR spectrum model with parameters $\vec\theta$. 
The processes run over hadronic and leptonic components, with $I_j(\vec\theta)$ including  CR protons, helium, electrons and positrons. 
 $ M_{jk}$ can include the dispersion in the \Fermi-LAT \gray energy measurements.
 The matrix $ M_{jk}$ is pre-computed for a given binning of data and model, so that the response can be computed very fast for a large number of model parameters.
The method is general and can be extended to include  cosmic-ray direct measurements, synchrotron emissivities etc.
Once the matrices and data are defined, the problem is a purely mathematical one.

\subsection{Energy dispersion}

The \Fermi-LAT energy measurement has a precision of about 15\%, becoming worse at low energies.
Accounting for the energy dispersion is especially important below 200 MeV, where photons are both lost from the measured range
and gained from higher and lower energies, depending on the input spectrum.
The energy dispersion is defined as $p(\Emeas|\Etrue)$, where $\Etrue,\Emeas$ are the true and measured \gray energies respectively.
 $p(\Emeas|\Etrue)$ is obtained in matrix form using the \Fermi-LAT Science Tools, 
 for the same event selection and response as used to derive the emissivities. 
The approximation is made that the response averaged over the full sky  is appropriate; the variations are small and  average out making this a good approximation.
The measured spectrum is also weighted by the \Fermi-LAT effective area which is a function of energy, varying rapidly at low energies. 
The full transfer function from the model proton and helium spectra to measured emissivities in energy bands can be expressed as a single matrix, allowing fast computation for statistical analysis.
\section{Bayesian analysis} 
The likelihood function is 
$L(data|\vec\theta)=\exp(-\Sigma_k(\epsilon_k^{pred(\vec\theta)}-\epsilon_k^{obs})^2/2\sigma_k^2)$
where $\epsilon_k^{obs}$ is the observed emissivity and $\sigma_k$ is the error estimate for the k'th energy bin.
By Bayes theorem the posterior probability of the model is 
$P(\vec\theta|data)={ P(\vec\theta) L(data|\vec\theta) \over P(data)}$
 where $P(\vec\theta)$ is the prior probability for the parameters $\vec\theta$ defining the model, and $P(data)$ is a normalizing factor (known as the {\it evidence} ) such that
$\int P(\vec\theta |data) d^N\vec\theta=1$.
Given the joint probability distribution of the parameters, we can  marginalize to obtain the distribution of any subset, including single parameters:
$P(\theta_i)=\int_{\theta\neq\theta_i} P(\vec\theta |data) d^{N-1}\vec\theta$.
Mean and standard deviations are
$\bar\theta_i=\int_{\theta\neq\theta_i}  \theta_i  P(\vec\theta |data) d^{N-1}\vec\theta$,
$\sigma(\theta_i)=\sqrt(\int_{\theta\neq\theta_i}  (\theta_i - \bar\theta_i)^2  P(\vec\theta |data) d^{N-1}\vec\theta)$. 
Since we are actually more interested in the CR spectrum than in a particular set of parameters, which are anyway highly correlated,
it is useful also to compute the mean and standard deviation of the gridded CR spectra $n_j=n(p_j)$:

$\bar n_j= \int n_j(\vec\theta)  P(\vec\theta |data) d^N\vec\theta$,
$\sigma (n_j)=  \sqrt(\int(n_j(\vec\theta)-\bar n_j)^2  P(\vec\theta |data) d^N\vec\theta)$. 
This defines a band containing the range of CR spectra.
This method  is therefore  a deconvolution using a basis of parameterized spectra\footnote{An alternative method, in principle preferable, would be to perform a parameter-free deconvolution of the CR spectra, with some regularizing prior; the present method is an intermediate approach.}.
The range of emissivities corresponding to this range can then be computed  (for each process) for comparison with the observed values, to illustrate the uncertainties.

In our previous analysis  \cite{2013arXiv1307.0497D,2013arXiv1303.6482D} we used an explicit parameter scan, which was limiting due to the large number of parameters.
The new analysis uses the MultiNest software\footnote{available at http://ccpforge.cse.rl.ac.uk/gf/project/multinest}
 \cite{2009MNRAS.398.1601F}, an advanced Bayesian package for multi-parameter fitting; among its advantages are that it does not require specification of a step-size, has handling of multi-modal posteriors, and allow computation of the statistics of functions of the parameters. We found it fast and reliable for the problem at hand.
The ranges of the parameters have to be specified.
As described below, we fit to eq.(\ref{fitfunction}), and there are a total of 10 parameters to be fitted.
The parameters in this application have rather limited ranges of reasonable values, and a flat prior over a prescribed range is  sufficient for the present purpose; more complex priors are straightforward to include if required.
MultiNest outputs full parameter chains,  mean and standard deviations of each parameter, and mean and standard deviations of  user-defined functions of the parameters, here  chosen to be the formula representing the CR spectrum as specified above.

\section{\gray production functions}



Recent reviews of hadronic \gray production can be found in  \cite{2012PhRvD..86d3004K,2013arXiv1307.0497D,2013arXiv1303.6482D,2015arXiv150505757D,2014ApJ...789..136K}.
Here we use the QGSJET-II-04 model \cite{2013EPJWC..5202001O,2012PhRvD..86d3004K}.
In addition to photon production in proton-proton collisions, we have to
account for the contribution to the photon yield by proton-helium, 
helium-proton, and helium-helium interactions. Since QGSJET-II includes
nuclei-nuclei interactions, the parametrization of \cite{2012PhRvD..86d3004K}
 can be used directly for all four reaction channels above the
transition energy.
For $A>4$ we apply a nuclear enhancement factor 
following \cite{2014ApJ...789..136K}. 
For proton energies below 20 GeV we use the \gray production functions described in 
\cite{2013arXiv1307.0497D,2013arXiv1303.6482D} 
;
these are for proton-proton collisions only, so we again apply the factor given in  \cite{2014ApJ...789..136K} for p-He, He-p, He-He and heavier nuclei.
The  helium abundance of interstellar gas  is 0.1 by number, and the interstellar gas composition for heavier nuclei uses \cite{2003ApJ...591.1220L}.
CR helium is taken into account explicitly, and
the CR composition for heavier nuclei is based on ACE and CREAM.
The contribution from CR and ISM with A$>$4 is about 10\% relative to A$\le$4, and is dominated by   CR, much less coming from heavy gas nuclei.

Bremsstrahlung from CR leptons is computed using the corresponding GALPROP routine.

\section{Synchrotron and direct constraints on electrons}
Synchrotron emission from electrons and positrons in the interstellar magnetic field  provides essential constraints on interstellar leptons, independent of
solar modulation. For a survey of experimental data and theoretical arguments see \cite{2011A&A...534A..54S,2011MNRAS.416.1152J,2013MNRAS.436.2127O}. 
We are not concerned here with the injection spectrum of electrons, or how the interstellar spectrum is affected by propagation, but just the
observational information on the interstellar spectrum.
As found in \cite{2011A&A...534A..54S}   the synchrotron brightness temperature spectral index $\beta$ changes from about 2.5  to 3 in the range from 100 MHz to several GHz. 
The relation of $\beta$ to the electron index $\alpha$ is $\beta=2+(\alpha-1)/2$,
so this corresponds to a steepening of the interstellar electron spectrum from index 2 to 3 at a few GeV electron energy. 
At high energies and frequencies, this is consistent with the Fermi-LAT direct measurements of electrons + positrons, which give an index $3.08\pm0.05$  \cite{2010PhRvD..82i2004A_short} and for electrons an index $3.19\pm0.07$ \cite{2012PhRvL.108a1103A_short}, the electrons fully dominating at low energies where bremsstrahlung is important;
this is also consistent with microwave synchrotron emission measured with {\it Planck}  \cite{2015arXiv150606660P}. 
Here solar modulation is negligible so that this consistency is a requirement, assuming the local measurement is typical of the interstellar medium near the solar position in the Galaxy. 
At low energies, only synchrotron is a reliable tracer of the interstellar electron spectrum,  due to the very large solar modulation (factor or 10 or more).
The relation of synchrotron to the electron spectrum is a complicated function \cite{2011A&A...534A..54S}, but it is useful to write a simplified version to illustrate the nature of the constraints on the energy of the electron spectral index break.
We use the formula from sec 2.1.1 and 2.3 of \cite{2011A&A...534A..54S}   and the synchrotron data from that paper.
We have for the synchrotron peak frequency for a electron energy E:
$\nu_{max}=0.29 \nu_c= 0.29 {B\over7.5\ \mu G}\  [{E\over GeV}]^2 \times 240$ MHz.
so 
$E = \sqrt({\nu_{max}\over 240\ MHz} / {0.29 B\over7.5\ \mu G})\ GeV$.
Note that E depends on $\sqrt\nu_{max}$ and $\sqrt B$ so is robust against uncertainties in these quantities.
The observed break $\nu$ is in the range 500--5000 MHz, B is between 5 and 10 $\mu G$, using a broad conservative range.
Hence taking extremes of $\nu/B$, $2 < E_{break} < 10$ GeV.
A narrower range is also reasonable: $\nu= 500-2000$ MHz, $B=5-8\ \mu G$, giving
$2.6 < E_{break} < 6.6$ GeV.
The upper limit is consistent with Fermi-LAT electrons which show no break down to 7 GeV.
However other experiments (AMS01, PAMELA) show the break must be below 3 GeV, since there is no break observed above this energy and
  solar modulation has the effect of enhancing the break in the directly measured spectrum.
Hence the adopted   break energy constraint from synchrotron and direct measurements   is 1--2.5 GeV. 
The synchrotron constraint is essential for the low end and the index below the break.
Since we use a smooth function for the break, the relation between the break parameter and the actual spectrum is not exact.
We use the \Fermi-LAT electron spectrum as measured for the spectrum above the break, including the Fermi-LAT normalization at 100 GeV and index range 3.1-3.2. 
Below the break, the experimental range of $\beta$ is 2.4 to 2.6, so from the above formula we use the range $\alpha=1.8-2.2$ as the constraint from synchrotron.


\section{Parameterization of spectra}
As explained in \cite{2013arXiv1307.0497D,2013arXiv1303.6482D}  we use a spectrum in the form of density per unit momentum $n(p)$, since this is expected to have a power-law shape in the theory of diffusive shock acceleration,
and hence serves as a suitable basis form on which propagation and other effects are superimposed.
The local emissivities require a turnover in the proton spectrum towards low energies.
A possible parameterization of CR spectra has a single sharp break in the spectral index, and this is often used; it was used in for example in \cite{2013arXiv1307.0497D,2013arXiv1303.6482D}.
 However this is unphysical, and a smoother function is more appropriate.
Hence smooth breaks in protons, helium and electrons are modelled;
the functional form used is 
$n(p)\propto 1\ /\ [ ({p\over p_{br}})^{\alpha_1/\delta}\ +\ ({p\over p_{br}})^{\alpha_2/\delta}]^\delta$
where $\alpha_1,\alpha_2$  are the indices below and above the break respectively for $\alpha_1<\alpha_2$,  and the spectrum breaks around a momentum  centred on $p_{br}$;
the parameter $\delta$ controls the sharpness of the break: smaller  $\delta$ produces a sharper break, typical values are   $\delta=$0.5 -- 1.5.
 It converges to the given power-laws at low and high $p$, with a smooth transition.
This is the same form as used for CR in supernova remnants in \cite{2013Sci...339..807A_short}, written in a symmetrical form in the indices for clarity.
This function has one more  parameter ($\delta$) than for a perfectly sharp break; it simply makes the break smoother in a controlled way.
Given a reference value $n_{ref}$ at $p_{ref}$ the normalized spectrum is:

\begin{equation}
n(p)= n_{ref}\times
[{({p_{ref}\over p_{br}})^{\alpha_1/\delta}\ +\ ({p_{ref} \over p_{br}})^{\alpha_2/\delta}]^\delta\ / \
 [({p\over p_{br}})^{\alpha_1/\delta}\ +\ ({p\over p_{br}})^{\alpha_2/\delta}}]^\delta
\label{fitfunction}
\end{equation}

The normalization $n_{ref}(p_{ref})$ is treated as a free parameter;  there are thus 5 parameters  each for  protons and leptons, making 10 parameters in total.
Since  gamma rays cannot distinguish an origin in  CR protons from CR helium, some assumptions have to be made to break the degeneracy.
 The ratio of protons and helium is fixed to that measured by PAMELA at 100 GeV, where solar modulation is absent while the  experimental error is  small. 
We take a reference proton momentum of 100 GeV, with a wide prior range since the normalization is to be determined from the gamma-ray data only.
The lepton (electron + positron) spectrum is only marginally constrained by gamma rays, so we rely on direct measurements for the normalization and high-energy spectral index, allowing a range with uniform prior.
We use the \Fermi-LAT lepton flux at 20 GeV \cite{2010PhRvD..82i2004A_short}.
For the low-energy (below about 1 GeV) spectral index we rely on synchrotron radiation constraints as described later. 
The prior range for the break position is based on both synchrotron and direct measurements.

\section{Results and Discussion}

\begin{table*}    
\label{models_summary}      
\caption{Summary of model fits to equation 6.1. 
Entries are prior range, posterior mean and standard deviation.
The proton parameters are constrained by the \gray emissivities, while the lepton parameters reflect mainly the prior from synchrotron and direct measurements. 
The parameters are highly correlated and degenerate, so the resulting spectrum derived from the full posterior (Fig 1) is preferred to the individual parameters.
The CR  density $n_{ref}$ is multiplied by $(c/4\pi)$ to give a flux in the usual units quoted in experiments.  $p_{ref}$= $10^5$ MeV for protons,  $2\times 10^4$ MeV for leptons.
}
\centering                          
\begin{tabular}{l l l l l l l l l l l l}        
\hline
Parameter&range: min&max&mean&std&units\\
\hline
Protons\\
\hline     
$(c/4\pi) n_{ref}$  &$1\times10^{-9}$ &$20\times10^{-9}$ &$6.4\times10^{-9}$ &$0.3\times10^{-9}$ & \fluxunits      \\         
$\alpha_1$ &2.2 &2.7 &2.37 & 0.09                   \\ 
$\alpha_2$ &2.6 &3.5 &2.82 & 0.05                   \\ 
$\delta$   &0.05 &1.0   &0.5 & 0.1                   \\ 
$p_{br}$   &1000&10000 &5870&2200 & MeV             \\
\hline
Leptons\\
\hline    
$(c/4\pi) n_{ref}$  &$1\times10^{-9}$ &$3\times10^{-9}$  &$2.2\times10^{-9}$ &$0.5\times10^{-9}$   & \fluxunits   \\          
$\alpha_1$ &1.8&2.2 &2.0 & 0.1                   \\ 
$\alpha_2$ &3.1&3.2 &3.15 & 0.03                   \\ 
$\delta$   &0.05&1   &0.47 & 0.25                   \\ 
$p_{br}$   &500&2000 &1130&4067 & MeV             \\
\hline   
\end{tabular}
\end{table*}

We illustrate the results using the raw emissivities accounting for energy dispersion; using the corrected values without applying dispersion leads to very similar results.
The fit results are given in Table 1 and Fig 1.
The parameters are highly correlated so the individual values are just indicative, while the resulting spectrum error band (Fig 1) uses the full posterior in all parameters.

\begin{figure*}
\centering
\resizebox{0.48\hsize}{!}{\includegraphics{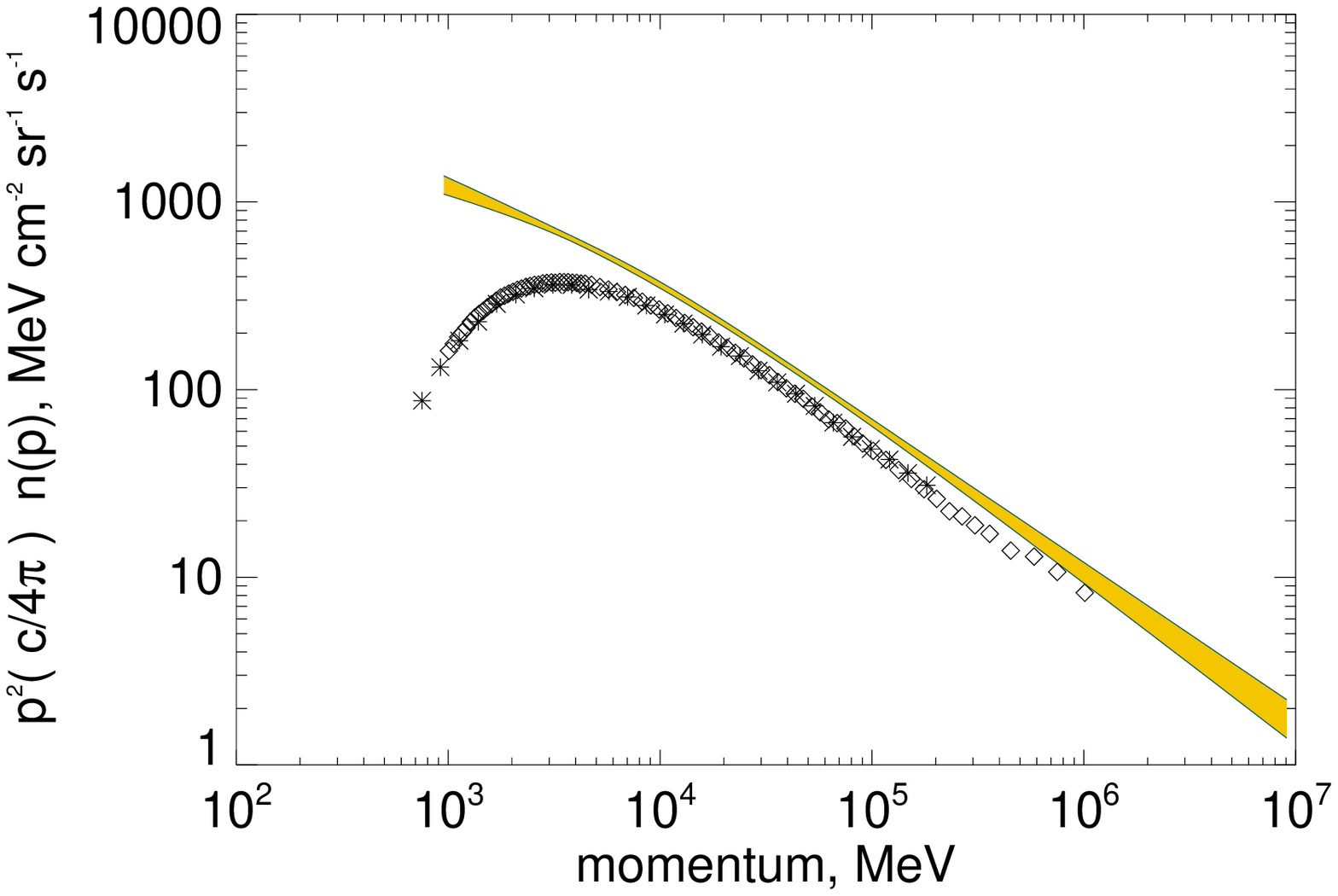}}
\resizebox{0.48\hsize}{!}{\includegraphics{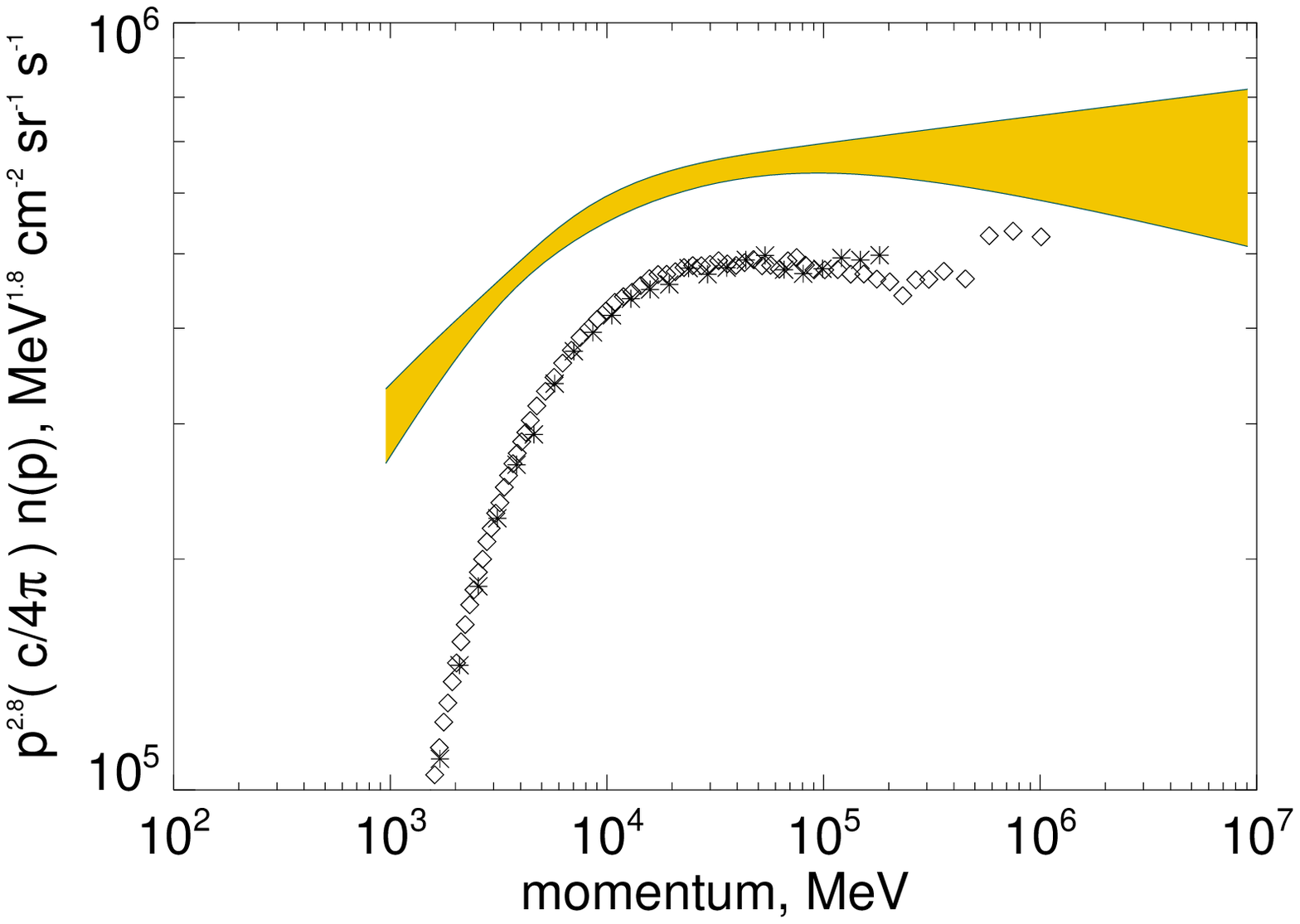}}
\resizebox{0.48\hsize}{!}{\includegraphics{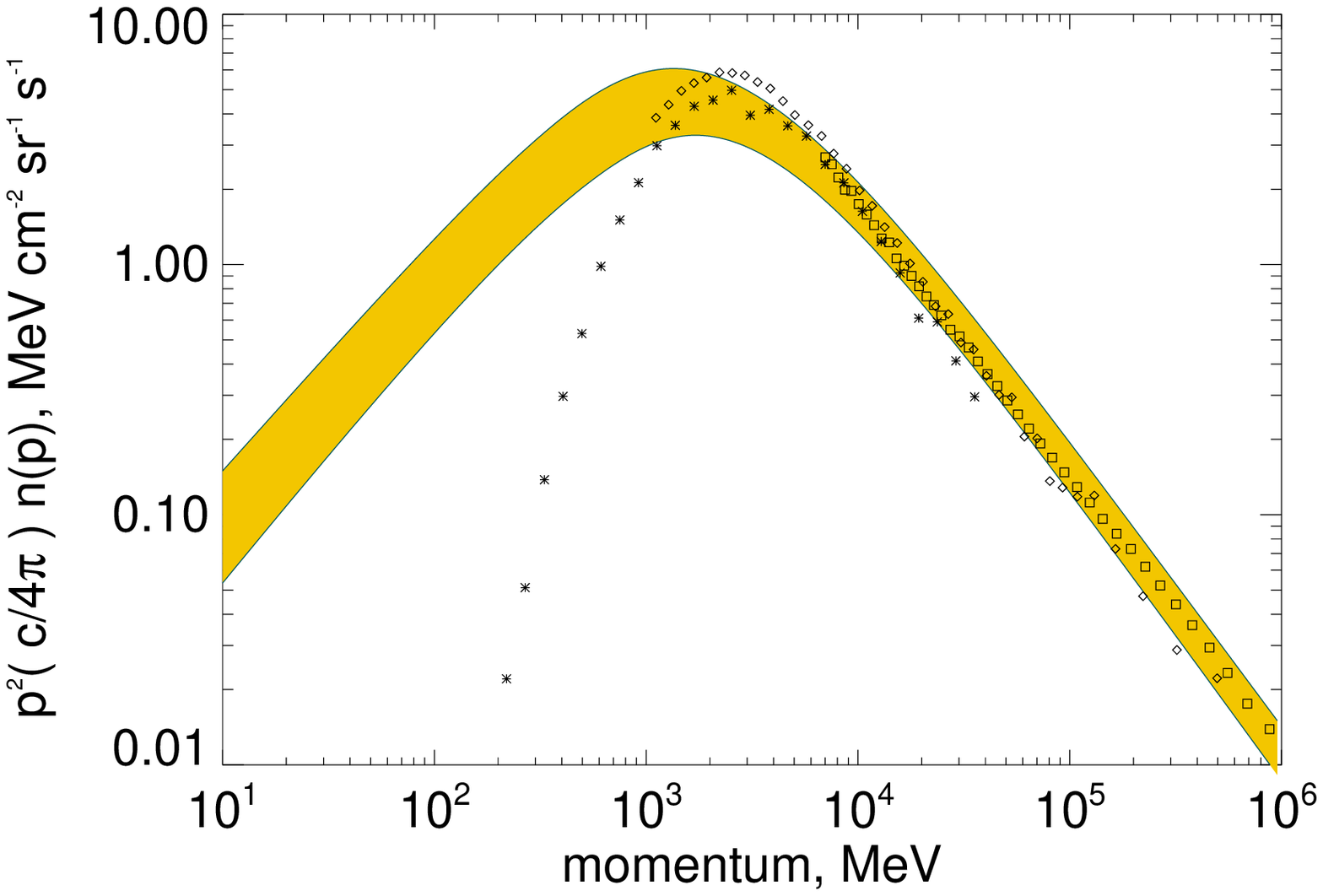}}
\resizebox{0.48\hsize}{!}{\includegraphics{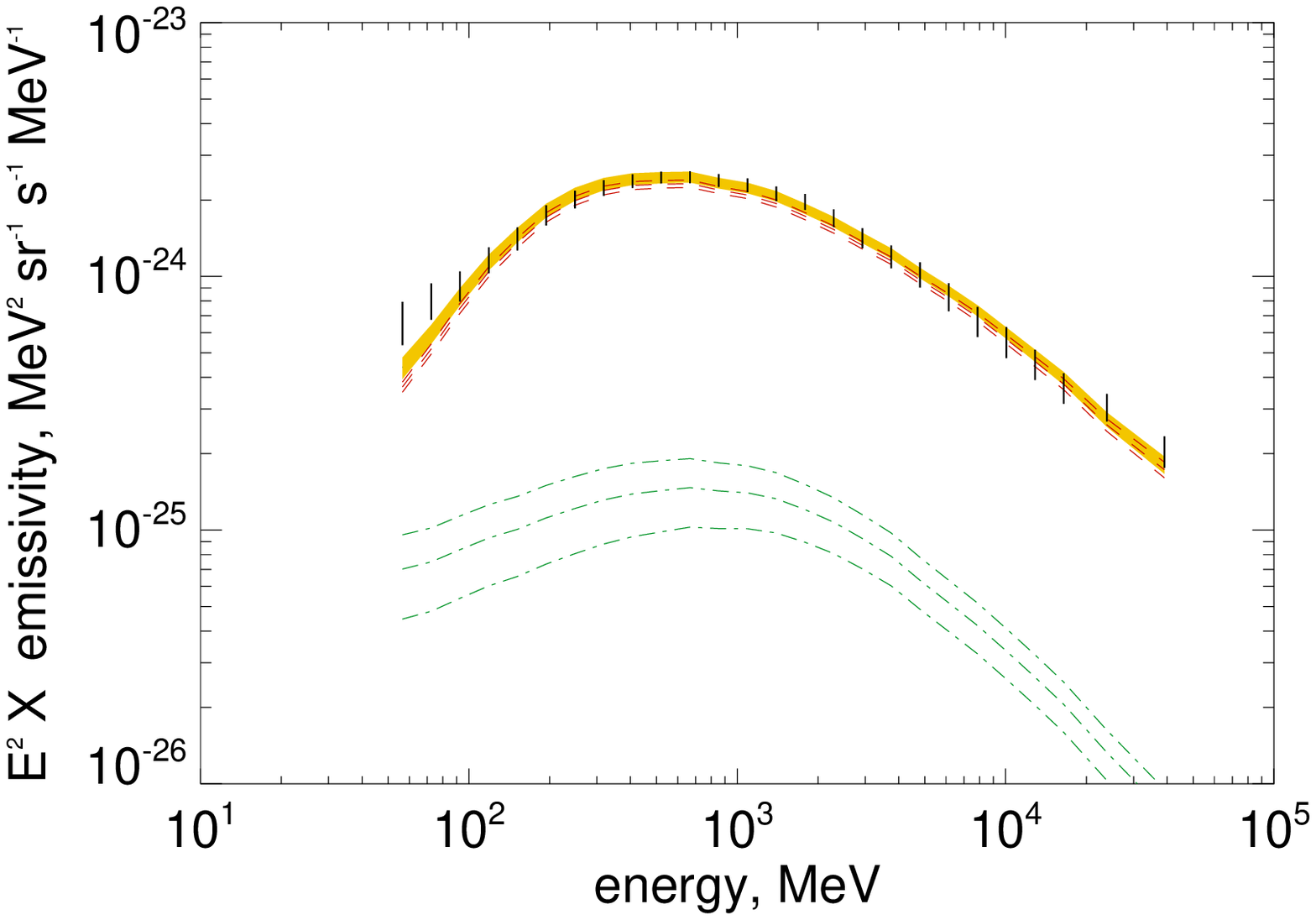}}
{\bf \color{red} PRELIMINARY}
\caption{Cosmic-ray and emissivity spectra derived from model fitting.
Yellow band shows model range. Model ranges are  1 standard deviation on the parameterized synthetic spectra.
Upper: Measured and derived cosmic-ray proton spectra. Data are AMS01 (asterisks) and PAMELA (diamonds). 
Spectra are multiplied by  $p^2$, and also times $p^{2.8}$ to better show  the break.
Lower left: Measured and derived cosmic-ray electron spectra. Data are AMS01 (asterisks), PAMELA (diamonds), and \Fermi-LAT (squares). 
The CR  density $n(p)$ is multiplied by $(c/4\pi)$ to give a flux at high momenta in the usual units quoted in experiments.
Lower right: $Fermi$-LAT emissivity data (vertical bars) and  model, with red and green curves showing the hadronic and leptonic  bremsstrahlung contributions; the yellow band
 shows the total. 
}
\label{results1}
\end{figure*}

The proton spectrum spectral index above the break is recovered in good agreement with direct measurements (beyond the effects of solar modulation); note that it has been determined from the gamma-ray data alone.
The spectral break is significant, the index being smaller by $\sim0.5$ below the break; this is where solar modulation affects the direct measurements, as can clearly be seen in the difference between PAMELA, AMS01 and the interstellar spectrum. 
Note that we address a break in the {\it momentum} spectrum, not kinetic energy; the latter representation is not appropriate, see \cite{2013arXiv1307.0497D,2013arXiv1303.6482D}.
The overall normalization of the proton spectrum is about 30\% larger than direct measurements, even in the high-energy region free from modulation.
This can be a real physical effect since there is no guarantee that the direct measurements represent the average in the local region within one kpc probed by the gamma rays.
There are spectral variations between the local region and the rest of the Galaxy \cite{2015PhRvD..91h3012G,2015arXiv150507601N}, and the region probed by the emissivity analysis might be not so ``local'' in this sense.
However there are sufficient uncertainties in the analysis to make such an interpretation  premature at this stage: hadronic cross-sections have still significant uncertainties especially for CR and target nuclei with $A>1$.
Uncertainties in the gas column densities including gas not traced by HI or CO are still significant, even though the emissivity derivation is for atomic hydrogen well traced by the 21-cm line, correlations between the gas phases does not allow full separation of the components. Ionized hydrogen is accounted for in the analysis, but again cannot necessarily be fully separated from the other components.
We note that  \cite{2015arXiv150600047C} implicitly finds a similar trend in the normalization: that analysis included proton and helium direct measurements in the fitting so is more constrained to agree with those. 
In fact their proton spectrum is slightly above direct measurements, and their predicted emissivities are slightly below the measurements; so to reproduce the emissivities the proton spectrum would have to be increased further relative to direct measurements, in accord to what we find here. Different cross-sections and analysis techniques (see below) can account for any remaining differences.
  One important use of the interstellar spectrum is for solar modulation studies; in this case it would be appropriate to use the spectral shape, which is well determined, but renormalize the spectrum to agree with direct measurements at high energies. This would attribute all the normalization difference to the factors mentioned above.

The interpretation of the curvature in the proton spectrum is beyond the present work, but it is generally consistent with expectation based on CR propagation implied by the B/C ratio, which peaks at a few GeV, implying either a break in the diffusion coefficient, effects of diffusive reacceleration, or convective transport. 
Any of these effects will cause a low-energy break in the primary spectra including that of protons, superimposed on a power-law injection spectrum \cite{2007ARNPS..57..285S}.
Note the apparent difference in index between the local proton spectrum and the large-scale Galaxy \cite{2015PhRvD..91h3012G,2015arXiv150507601N}; the latter spectrum is evidently harder. 
The origin of the difference is not clear at present, but it is interesting that the index from local gamma rays and local direct measurements are consistent.
There are further suggestions of local CR spectral variations \cite{2015arXiv150406472K}.

\section{Comparison of analyses}
The present analysis differs from that of  \cite{2015arXiv150600047C} (Cas15) in various ways; both approaches are valid and are complementary, emphasizing different aspects of the subject.
The main differences can be summarized as follows:
Cas15 performs an iterative correction for energy dispersion, we include it in the response to the raw emissivities.
Cas15 includes direct measurements of protons and helium in the fit, combined with the emissivities. We use only \gray data for the hadronic contribution, and adopt a He/p ratio from direct measurements, assuming equal spectral shapes for p and He.
Cas15 uses a velocity term to obtain the low-energy turnover in the proton spectrum; we use a smoothly broken power law for more flexibility. 
Cas15 uses a flux spectrum in kinetic energy, we use a density spectrum in momentum.
Cas15 uses cross-sections from Kamae; we use a combination from Dermer and QGSJET-II.
Cas15 uses a lepton spectrum from direct measurements combined with the \gray emissivities, we combine direct measurements with constraints from synchrotron.
Cas15 uses Minuit for the fitting, we use a Bayesian scheme with MultiNest.



\acknowledgments{
I thank Jean-Marc Casandjian for providing the emissivity data and useful discussions, and Chuck Dermer, Michael Kachelriess and Sergey Ostapchenko for their hadronic  production software and many  discussions about the physical processes.

The \textit{Fermi}-LAT Collaboration acknowledges support for LAT development, operation and data analysis from NASA and DOE (United States), CEA/Irfu and IN2P3/CNRS (France), ASI and INFN (Italy), MEXT, KEK, and JAXA (Japan), and the K.A.~Wallenberg Foundation, the Swedish Research Council and the National Space Board (Sweden). Science analysis support in the operations phase from INAF (Italy) and CNES (France) is also gratefully acknowledged.}

\bibliographystyle{JHEP}

\bibliography{sourcepop_notitles,luminosity_notitles,strong_notitles,CRemiss_notitles,andy_notitles.bib} 

\end{document}